\documentclass[aps,prb,a4paper,preprint,showpacs]{revtex4}
\usepackage{amsmath}
\usepackage{amsfonts}
\usepackage{amssymb}
\usepackage{slashbox}
\usepackage{graphicx}
\usepackage{amsbsy}

\begin{document}
\title{Shot noise of the conductance through a superconducting barrier in graphene }
\author{Mi Liu and Rui Zhu\renewcommand{\thefootnote}{*}\footnote{Corresponding author.
Electronic address:
rzhu@scut.edu.cn} }
\address{Department of Physics, South China University of Technology,
Guangzhou 510641, People's Republic of China }

\begin{abstract}

We investigated the conductance and shot noise properties of quasiparticle-transport through a superconducting barrier in graphene. Based on the Blonder, Tinkham, and Klapwijk (BTK) formulation, the theory to investigate the transport properties in the superconductive graphene is developed. In comparison, we considered the two cases that are the transport in the presence and absence of the specular Andreev reflection. It is shown that the conductance and shot noise exhibit essentially different features in the two cases. It is found that the shot noise is suppressed as a result of more tunneling channels contributing to the transport when the superconducting gate is applied. The dependence of the shot noise behavior on the potential strength and the width of the superconducting barrier differs in the two cases. In the presence of the specular Andreev reflection, the shot noise spectrum is more sensitive to the potential strength and the width of the superconducting barrier. In both of the two cases, total transmission occurs at a certain parameter setting, which contributes greatly to the conductance and suppresses the shot noise at the same time.

\end{abstract}

\pacs {74.20.-z, 72.70.+m, 72.10.-d}

\maketitle

\narrowtext

\section{Introduction}

In the past decade, there has been a great deal of interest in studying the physical properties of graphene both theoretically and experimentally\cite{BeenakkerPRL2006,BeenakkerRMP2008,Castro NetoRMP2009}. In graphene, the low-energy excitations are massless and chiral Dirac fermions with linear energy dispersion near the Dirac point. Because of the unique energy dispersion in graphene, there are some special phenomena such as the unconventional quantum Hall effect, Klein paradox, etc\cite{Castro NetoRMP2009,BaiPRB2009,BaiPRB2007,BaiSuperMicros2011,KatsnelsonNature2006}. Particularly relevant to the present work, the presence of the Dirac Fermions in the graphene-based superconductor junctions results in the specular Andreev reflection\cite{BeenakkerPRL2006,BeenakkerRMP2008}.

The Andreev reflection\cite{AndreevJETP1964} describes the tunneling phenomenon of electron excitation converting into hole excitation by the superconducting pair potential. The interface between a metal and a superconductor can reflect a negatively charged electron incident from the metal into a positively charged hole, while the missing charge enters the superconductor forming a Cooper pair. Usually the hole is reflected back along the path of the incident electron in the conventional materials, which is called ``retro-Andreev reflection''. However, the Andreev reflection in undoped graphene is specular---the so-called ''specular Andreev reflection'', in which the reflected angle is inverted\cite{BeenakkerPRL2006,BeenakkerRMP2008,BaiSuperMicros2011,BenjaminPRB2008}. Because the graphene needs to be described by the Dirac-like equation rather than the usual Schrodinger equation, the superconductive graphene needs to be described by the Dirac--Bogoliubov--de Gennes equation rather than the Bogoliubov--de Gennes equation for usual superconductors\cite{BeenakkerPRL2006,BeenakkerRMP2008,BhattacharjeePRL2006,BaiPLA2010,BaiPLA2008,BaiAPL2008}. However, graphene is not a natural superconductor. Recent research has shown that superconductivity can be induced in a graphene layer in the presence of a superconducting electrode by means of the proximity effect\cite{BeenakkerPRL2006,TitovPRB2006,BhattacharjeePRL2006}. So far, the physical properties of the graphene-based superconductor junctions have been extensively studied and many important results have been obtained. Most of them focus on the transport properties of the graphene/superconductive graphene interface and only the conductance is considered\cite{BeenakkerPRL2006,BaiSuperMicros2011,BhattacharjeePRL2006,BaiPLA2010,JongPRB1994,BeenakkerPRB1992,
BlonderPRB1982}, which prompts us to investigate the transport properties through a superconducting barrier in graphene and especially focus on the shot noise spectrum.
In this work, based on the Dirac--Bogoliubov--de Gennes equation and the Blonder, Tinkham, and Klapwijk (BTK) formulation, we derive the shot noise formula and provide numerical results of the transport properties in the graphene/superconductive graphene/graphene (G/SG/G) heterostructure. Dependence of the shot noise spectrum on the structure parameters and its physical mechanisms are discussed.

\section{Theory and Model}
 We consider transmission through the G/SG/G heterostructure occupying the $x$-$y$ plane, the schematic of which is shown in Fig. 1. The growth direction of the graphene is taken to be the $x$-axis. The left G region extends from $x=-\infty$ to $x=0$ and the right G region extends from $x=a$ to $x=+\infty$. The superconductive region occupies the $0<x<a$ region. The electron and hole excitations are described by the Dirac--Bogoliubov--de Gennes equation\cite{BeenakkerPRL2006,BaiSuperMicros2011,BhattacharjeePRL2006,BaiPLA2010,JongPRB1994,BeenakkerPRB1992,
 BlonderPRB1982}:
 \begin{equation}
 \left( {\begin{array}{*{20}{c}}
   {H_\alpha-E_F} & {\Delta\left( r\right)}  \\
   {\Delta{^*\left( r\right)}} & {E_F-H_\alpha}  \\
\end{array}} \right)\Psi_\alpha = E\Psi_\alpha.
\label{DBG-equation}
\end{equation}
Here, $\Psi_\alpha=\left( \psi_{A\alpha},\psi_{B\alpha},\psi_{A\bar{\alpha}}^*,-\psi_{B\bar{\alpha}}^* \right)$ are the four-component wave functions for the electron and hole spinors. The index $\alpha$ denotes $K$ or $K'$ for the electrons or holes near the Dirac $K$ and $K'$ points. $\bar{\alpha}$ takes values $K'\left( K \right)$ for $\alpha=K\left( {K'} \right)$. $E_F$ denotes the Fermi energy. $A$ and $B$ denote the two inequivalent sites in the hexagonal lattice of graphene. The Hamiltonian $H_\alpha$ is given by
\begin{equation}
H_\alpha = -i\hbar v_F\left[ \sigma_x\partial_x + \rm{sgn} \left(\alpha \right)\sigma_y\partial_y \right]+U \left( r \right),
\end{equation}
where $v_F$ denotes the Fermi velocity of the quasiparticles in graphene and $\rm{sgn} \left( \alpha \right)$ takes the values of $\pm 1$ for $\alpha=K\left( K ' \right)$.

For $0<x<a$, the superconducting electrode on the top of the graphene layer induces a nonzero pair potential $\Delta\left( \bf{r} \right)$ via the proximity effect. We model the pair potential as
\begin{equation}
\Delta\left( \bf{r} \right) = \{{\begin{array}{l}
 \ 0, {\kern 50pt}  \rm{others},  \\
 \ \Delta_0 e^{i\phi}, {\kern 20pt}  0 < x < a,  \\
 \end{array}}
\end{equation}
where $\Delta_0$ and $\phi$ are the amplitude and the phase of the induced superconductive order parameter. The electrostatic potential $U\left( \bf{r} \right)$ in the G and SG regions can be tuned independently by a gate voltage or by doping. We take
\begin{equation}
U\left( \bf{r} \right) = \{{\begin{array}{l}
\ 0, {\kern 50pt}  \rm{others},  \\
\ -U_0, {\kern 25pt}  0 < x < a.  \\
 \end{array}}
\end{equation}

Eq. (\ref{DBG-equation}) can be solved straightforwardly to yield the wave functions $\Psi$ in the G and SG regions, respectively. In the G region, for the electrons and holes traveling in the $\pm x$ directions with a transverse momentum $k_y=q$ and energy $\varepsilon$, the wave functions are given by
\begin{equation}
\begin{array}{l}
 \begin{array}{*{20}c}
  { \Psi _N ^{e\pm}  = \frac{exp\left( {iqy\pm ikx} \right)}{\sqrt{\cos \alpha}}\left({ \begin{array}{l}
 \begin{array}{*{20}c}
 {exp\left( {\mp i\alpha/2} \right)}\\
   \end{array}\\
   \begin{array}{*{20}c}
   {\pm exp\left( {\pm i\alpha/2} \right)}\\
   \end{array}\\
   \begin{array}{*{20}c}
   {\kern 30pt}{0}\\
   \end{array}\\
   \begin{array}{*{20}c}
   {\kern 30pt}{0}\\
   \end{array}\\
   \end{array}} \right),}\\
\end{array}\\
 \begin{array}{*{20}c}
  { \Psi _N ^{h\pm}  = \frac{exp\left( {iqy\pm ik'x} \right)}{\sqrt{\cos \alpha'}}\left( {\begin{array}{l}
 \begin{array}{*{20}c}
 {\kern 30pt}{0}\\
 \end{array}\\
   \begin{array}{*{20}c}
 {\kern 30pt}{0}\\
 \end{array}\\
   \begin{array}{*{20}c}
 {exp\left( {\mp i\alpha'/2} \right)}\\
 \end{array}\\
   \begin{array}{*{20}c}
 {\mp exp\left( {\pm i\alpha'/2} \right)}\\
 \end{array}\\
   \end{array}} \right),}\\
   \end{array}\\
 \end{array}
\end{equation}
with
\begin{equation}
\begin{array}{l}
 \begin{array}{*{20}c}
 {\sin \alpha  =\frac{\hbar v_F q}{\varepsilon + E_F}},\\
\end{array}\\
 \begin{array}{*{20}c}
 {\sin \alpha'  =\frac{\hbar v_F q}{\varepsilon - E_F}},\\
\end{array}\\
 \begin{array}{*{20}c}
 {k = \frac{\varepsilon + E_F}{\hbar v_F}\cos \alpha },\\
\end{array}\\
 \begin{array}{*{20}c}
 {k' = \frac{\varepsilon - E_F}{\hbar v_F}\cos \alpha' }.\\
\end{array}\\
 \end{array}
\end{equation}
$\alpha$ is the incident angle of the electron and $\alpha'$ is the reflection angle of the hole. Note that for the Andreev process to take place, the maximum incident angle of the electron is given by
\begin{equation}
\alpha_c =\arcsin\left( {\frac{\left| {\varepsilon - E_F} \right|}{\varepsilon + E_F}}\right).
\end{equation}

In the SG region, the quasiparticles are mixtures of the electrons and holes. The wave functions of these quasiparticles moving along the $\pm x$-direction with the transverse momentum $q$ and energy $\varepsilon$ has the form of
\begin{equation}
\begin{array}{l}
 \begin{array}{*{20}c}
  { \Psi _S ^{e\pm}  = e^{iqy\pm i\left( {k_0 - i\kappa} \right)}\left({\begin{array}{l}
 \begin{array}{*{20}c}
  {{\kern 10pt}exp\left( {-i\beta} \right)}\\
  \end{array}\\
 \begin{array}{*{20}c}
  {\pm exp\left( {-i\beta \pm i\gamma} \right)}\\
  \end{array}\\
 \begin{array}{*{20}c}
  {{\kern 10pt}exp\left( {-i\phi} \right)}\\
  \end{array}\\
 \begin{array}{*{20}c}
  {\pm exp\left({ \pm i\gamma -i\phi} \right)}\\
  \end{array}\\
 \end{array} }\right), }\\
\end{array}\\
 \begin{array}{*{20}c}
  { \Psi _S ^{h\pm}  = e^{iqy\mp i\left( {k_0 + i\kappa} \right)}\left({\begin{array}{l}
 \begin{array}{*{20}c}
  {{\kern 10pt}exp\left( {i\beta} \right)}\\
  \end{array}\\
 \begin{array}{*{20}c}
  {\mp exp\left( {i\beta \mp i\gamma} \right)}\\
  \end{array}\\
 \begin{array}{*{20}c}
  {{\kern 10pt}exp\left( {-i\phi} \right)}\\
  \end{array}\\
 \begin{array}{*{20}c}
  {\mp exp\left( {\mp i\gamma -i\phi} \right)}\\
  \end{array}\\
 \end{array} }\right). } \\
 \end{array}\\
 \end{array}
 \label{SGWaveFunctions}
\end{equation}
The parameters $\beta$, $\gamma$, $k_0$, and $\kappa$ are defined by
\begin{equation}
\begin{array}{l}
 \begin{array}{*{20}c}
{\beta =\{ \begin{array}{l}
 \arccos\left( {\varepsilon/\Delta_0} \right), {\kern 50pt}  \varepsilon < \Delta_0 ,\\
 -i arcosh\left( {\varepsilon/\Delta_0}\right), {\kern 33pt}  \varepsilon > \Delta_0 , \\
 \end{array}}\\
\end{array}\\
 \begin{array}{*{20}c}
 {\gamma =\arcsin\left[ {\hbar v_F q/\left( U_0 + E_F \right)}\right],}\\
\end{array}\\
 \begin{array}{*{20}c}
 {k_0 = \sqrt{\left( E_F + U_0 \right)^2/\left( \hbar v_F\right)^2 - q^2},}\\
\end{array}\\
 \begin{array}{*{20}c}
 {\kappa = \frac{\left( E_F +U_0 \right)\Delta_0}{\left( \hbar v_F \right)^2 k_0}\sin\beta ,}\\
\end{array}\\
 \end{array}
\end{equation}
Taking into account both the Andreev and normal reflection processes, the wave functions in the left G, SG, and right G regions can be written as
\begin{equation}
\begin{array}{l}
 \begin{array}{*{20}c}
 {\Psi_1 = \Psi_N^{e +}+r_c\Psi_N^{e -}+r_{Ac}\Psi_N^{h-},}
\end{array}\\
 \begin{array}{*{20}c}
 {\Psi_2 = A\Psi_S^{e +}+B\Psi_S^{e -}+C\Psi_S^{h+}+D\Psi_S^{h-},}\\
\end{array}\\
 \begin{array}{*{20}c}
 {\Psi_3 = t_c\Psi_N^{e +}+t_{Ac}\Psi_N^{h+},}\\
\end{array}\\
 \end{array}
 \label{WaveFunctions}
\end{equation}
respectively. Here, $r_c$ and $r_{Ac}$ are the amplitudes of the normal and Andreev reflections, respectively; $t_c$ and $t_{Ac}$ are the amplitudes of the normal and Andreev transmissions, respectively. $A$, $B$, $C$, and $D$ are the amplitudes of electronlike and holeslike quasiparticles in the SG region. All the amplitudes in Eq. (\ref{WaveFunctions}) can be determined by demanding wave function continuity at the interfaces. These boundary conditions are given by
\begin{equation}
\Psi_1\left( 0 \right)=\Psi_2\left( 0 \right), {\kern 10pt} \Psi_2\left( a \right)=\Psi_3\left( a \right).
\end{equation}
The scattering amplitudes can be obtained by numerically solving these continuity equations.

The electron and hole operators of the outgoing states are related to the electron and hole operators of the incoming states via the scattering matrix\cite{BlanterRep2000,MuzykantskiiPRB1994},
\begin{equation}
\left( {\begin{array}{*{20}{c}}
   {b_{Le}}  \\
   {b_{Re}}  \\
\end{array}} \right) = \left( {\begin{array}{*{20}{c}}
   {S_{LL}^{e e}} & {S_{RL}^{e e}}  \\
   {S_{LR}^{e e}} & {S_{RR}^{e e}}  \\
\end{array}} \right)\left( {\begin{array}{*{20}{c}}
   {a_{Le}}  \\
   {a_{Re}}  \\
\end{array}} \right),{\kern 10pt}\\
\left( {\begin{array}{*{20}{c}}
   {b_{Lh}}  \\
   {b_{Rh}}  \\
\end{array}} \right) = \left( {\begin{array}{*{20}{c}}
   {S_{LL}^{e h}} & {S_{RL}^{e h}}  \\
   {S_{LR}^{e h}} & {S_{RR}^{e h}}  \\
\end{array}} \right)\left( {\begin{array}{*{20}{c}}
   {a_{Lh}}  \\
   {a_{Rh}}  \\
\end{array}} \right),
\label{ScatteringMatrix}
\end{equation}
where the element $S^{e e}$ gives the outgoing electron current amplitude in response to an incoming electron current amplitude and $S^{e h}$ gives the outgoing hole current amplitude in response to an incoming electron current amplitude. The generalized current operator for the electrons and holes in the left electrode can be written as\cite{BlanterRep2000,MuzykantskiiPRB1994,AnantramPRB1996}
\begin{equation}
I_L\left( t \right)  = \frac{e}{2\pi \hbar}\int_0^\infty dEdE' e^{i\left( {E-E'}\right)t/\hbar}\left[ {\begin{array}{*{20}{c}}{<a_{Le}^+\left( E \right) a_{Le}\left( {E'} \right)> - <b_{Le}^+\left( E \right) b_{Le}\left( {E'} \right)>}\\ {+ <b_{Lh}^+\left( {-E} \right) b_{Lh}\left( {-E'} \right)>}\\
\end{array}}\right].
\end{equation}
Substituting the scattering matrix in Eq. (\ref{ScatteringMatrix}), we can obtain
\begin{equation}
I_\alpha \left( t \right)  = \frac{e}{2\pi \hbar}\sum_{\mu m\gamma p}\int_0^\infty dEdE' e^{i\left( {E-E'}\right)t/\hbar}\left[ {\begin{array}{*{20}{c}}{a_{\mu e m}^+\left( E \right) A_{\mu\gamma}^{m p}\left( {\alpha,E,E'}\right) a_{\gamma e p}\left( {E'}\right)}\\
{ + \sum_{n} a_{\mu h m}^+\left( E \right) S_{\alpha\mu n m}^h\left( {E'}\right) a_{\gamma h p}\left({E'} \right)}\\
\end{array}}\right],
\end{equation}
where
\begin{equation}
A_{\mu\gamma}^{m p}\left( {\alpha,E,E'}\right) = \delta_{\mu\alpha}\delta_{\gamma\alpha}\delta_{m n}\delta_{p n} - \sum_n S_{\alpha\mu n m}^{e+}\left( E \right) S_{\alpha\gamma n p}^e \left( {E'} \right).
\end{equation}

The general expression for the current fluctuations between contacts $\alpha$ and $\beta$ is
\begin{equation}
S_{\alpha\beta}\left( {t-t'} \right) = \frac{1}{2} <\Delta I_\alpha\left( t \right)\delta I_\beta\left( {t'} \right) + \delta I_\beta\left( {t'}\right)\Delta I_\alpha\left( t \right)>,
\end{equation}
with its Fourier transform
\begin{equation}
2\pi\delta\left( {\omega - \omega'} \right)S_{\alpha\beta}\left( \omega \right) =  <\Delta I_\alpha\left( \omega \right)\delta I_\beta\left( {\omega'} \right) + \delta I_\beta\left( {\omega'}\right)\Delta I_\alpha\left( \omega \right)> .
\end{equation}
We restrict our consideration to coherent tunneling and neglect the Coulomb interaction. In the zero-frequency limit with $\omega=0$, providing all the information above, we can express the noise power as
\begin{equation}
S_{\alpha\beta}=\frac{e^2}{2\pi\hbar}\sum_{\mu m \gamma p}\int_0^\infty dE\times \left[ {\begin{array}{*{20}{c}}
   {\begin{array}{*{20}{c}}{A_{\mu\gamma}^{m p}\left( {\alpha,E,E}\right) A_{\gamma\mu}^{p m}\left( {\beta,E,E}\right)}\\
   {\times{f_{\mu e}\left( E \right)\left[ {1-f_{\eta e}\left( E\right)} \right] + f_{\eta e}\left( E \right)\left[ {1-f_{\mu e}\left( E \right)} \right]}}\\
   \end{array} }  \\
  {\begin{array}{*{20}{c}}{+\sum_{nl} S_{\alpha\mu n m}^{h*}\left( E \right) S_{\alpha\gamma n p}^h\left( E\right) S_{\beta\gamma l p}^{h*}\left( E \right) S_{\beta\mu l m}^h\left( E \right)}\\
   {\times{f_{\mu e}\left( {-E} \right)\left[ {1-f_{\eta e}\left( {-E}\right)} \right] + f_{\eta e}\left( {-E} \right)\left[ {1-f_{\mu e}\left( {-E} \right)} \right]}}\\
 \end{array}   }  \\
\end{array}} \right].
\end{equation}
By introducing the distribution function for the electrons $f_e\left( E \right)=\left[ {exp\left[ {\left( {E-eV}\right)/k_B T}\right]+1}\right]^{-1}$ and that for the holes $f_h\left( E \right)=\left[ {exp\left[ {\left( {E+eV}\right)/k_B T}\right]+1}\right]^{-1}$, the zero-temperature conductance can be obtained from the current operator and from the usual quantum statistical assumptions for the averages and correlations of the electron and hole operators in the normal reservoirs as
\begin{equation}
G\left( {eV} \right) = G_0\int_0^{\alpha_c}\left( {1-\left| r_c \right|^2 + \left| r_{Ac}\right|^2\frac{\cos\alpha'}{\cos\alpha}} \right)\cos\alpha d\alpha .
\end{equation}
The shot noise power can be obtained as
\begin{equation}
S\left( 0 \right)=4eG_0\int_0^{\alpha_c}\left[ {\left|r_c \right|^2 \left|t_c \right|^2 + \left|r_{Ac} \right|^2 \left|t_{Ac} \right|^2\left( \frac{\cos\alpha'}{\cos\alpha} \right)^2}\right]\cos\alpha d\alpha ,
\end{equation}
where $G_0=4e^2 N\left( {eV}\right)/h$ is the ballistic conductance of metallic graphene, $V$ is the bias voltage, and $N\left( \varepsilon \right)= \left( {\varepsilon+E_F} \right) w/\left( {\pi\hbar v_F}\right)$ denotes the number of available channels for a graphene sample of width $w$.

\section{Numerical results and discussion}

Now we present numerical results for the Andreev reflection coefficients and the tunneling conductance for the $G/SG/G$ junction with $U_0\neq 0$. In this condition, there is a large mismatch of Fermi surfaces on the G and SG sides. Such a mismatch is well known to act as an effective barrier\cite{BaiSuperMicros2011,BhattacharjeePRL2006,BaiPLA2010}. The transport properties of the two cases of $E_F\gg \Delta_0$ ($E_F=0$) and $E_F\ll \Delta_0 $ are significantly different. In the case of $E_F\gg \Delta_0$, i.e., the incident electron and the reflected hole both lie in the conduction band, which results in the ``retro-Andreev reflection'', while in the case of $E_F\ll \Delta_0 $, only the ``specular Andreev reflection'' takes place, since the incident electron in the conduction band is converted into the reflected hole in the valence band\cite{BeenakkerPRL2006,BaiSuperMicros2011,BhattacharjeePRL2006,BaiPLA2010}. In general, it is very difficult to reach the regime $E_F\ll \Delta_0$ in experiment\cite{BeenakkerPRL2006}. So we only consider the case of $E_F\gg \Delta_0$ and the condition of comparable $E_F$ and $\Delta_0$, in the latter of which both normal Andreev reflection and specular Andreev reflection play roles and we can see that the retroreflection crosses over to specular Andreev reflection.

Firstly, we consider dependence of the transport properties on the SG-barrier thickness. The tunneling conductance through the SG-barrier as a function of the thickness for different potential strengths $U_0$ is shown in Fig. 2. Dimensionless thickness $k_0 a$ is used. The solid and dashed lines correspond to $U_0/E_F = 2$ and 10, respectively. When the bias voltage is small, the conductance exhibits oscillation features. Their oscillation amplitudes decay with increase of the thickness of the SG layer. The oscillation period of the shot noise power is the same as that of the conductance. The shot noise characterises correlations of the current. It can be seen from the curves that the growing trend of the shot noise is opposite to that of the conductance. When the conductance reaches the maximum value, the shot noise approaches the minimum value. This can be interpreted by the relation between the shot noise and the properties of the scatterer. A coherent conductor with all the transmission channels open (The open channel means that the transmission probability is close to one.) has minimal shot noise with the Fano factor approaching 0. A coherent conductor with all closed channels (The closed channel means that the transmission probability is close to zero.) has maximal shot noise with the Fano factor approaching 1. The strength of the shot noise is in the middle of the two limits when a conductor has open and closed channels coexistent. Therefore large transmission probabilities enhance the conductance and suppress the shot noise. It can also be seen in the panels (b) and (d) that for the case of $E_F=\Delta_0$ the conductance and shot noise decrease with the increase of the thickness of the SG layer. This is because that the parameter $\kappa$ is proportional to $\left( {\Delta/\hbar v_F}\right)\sin\beta$ for identical $U_0$ and it is in the exponential form of $e^{\pm\kappa a}$ in the wave function as shown in Eq. (\ref{SGWaveFunctions}). It is larger for the case of $E_F=\Delta_0$ than for the case of $E_F\gg\Delta_0$, which results in the quick decrease of the shot noise and the conductance. On the other hand, we can see from Fig. 2 that with the increase of the potential strength, the tunneling conductance decreased, which illustrates that the Fermi surface mismatch between the normal and superconducting regions suppresses transmission. We also considered the dependence of the tunneling conductance on the value of $U_0$ for small bias voltages. As expected, we found that the oscillation amplitude decreases monotonically with the increase of $U_0$ in the case of $E_F=\Delta_0$ and finally approaches a constant value.

In Figs. 3 and 4, we provide numerical results of the tunneling conductance as a function of the bias voltage $V$. Similar results to Beenakker's\cite{BeenakkerPRL2006} are obtained. In the limit of $V\rightarrow 0$, all the conductances have the same value of $4/3$. For the case of $\Delta_0 \geq E_F$, a sharp change in the conductance occurs at $eV = \Delta_0$ and all the conductances vanish at $eV=E_F$, which is shown in Fig. 3.  This is because of that no Andreev reflection occurs for all the incident angles (the critical angle of incidence $\alpha_c=0$) when $\varepsilon=E_F$. For the small bias voltages before the turning point $eV=E_F$, the conductance decreases with the increase of the SG energy gap. The conductance curves exhibit oscillatory behavior in the region of $eV>\Delta_0$, which is different from the condition considered by Beenakker. By analyzing the transmission coefficients of the system, we found that in the condition of $eV>\Delta_0$ the transmission spectrum demonstrates oscillatory behavior and the oscillating period increases with the increase of the SG energy gap. The oscillatory behavior originates from the effect of the quantum-mechanical interference between the electron-like and hole-like quasiparticles\cite{McMillanandPRL1966,PasanaiPhysica.C2014,DongPRB2003} in the SG barrier. This effect gives rise to oscillations in the reflection and transmission probabilities for the incident energies larger than the gap energy.

In Fig. 4, numerical results of the conductance for different SG-barrier thickness $a$ are provided. It can be seen that the conductance is sensitive to the SG-barrier thickness, especially in the condition of $eV>\Delta_0$. In this condition, the oscillation period increases and the oscillation amplitude decreases sharply with the increase of the thickness. This is also a result of the exponential term $e^{\pm\kappa a}$ in the wave function. It can be interpreted by the $ a \rightarrow \infty $ limit. In this limit, the components of the wave function with the exponential term $e^{+ \kappa a}$ are nonphysical, therefore only two of the components of wave function in the SG-region are physical. The model reduces to that of the the G/SG-junction, in other words, the model proposed by Beenakker\cite{BeenakkerPRL2006} is obtained.

We also considered the shot noise properties of the conductance through the G/SG/G structure. Numerical results of the shot noise and the Fano factor are provided in Fig. 5. The parameters of panels (a) and (c) are the same with Fig. 4 and teh parameters of (b) and (b) are the same with Fig. 3. We can see from the curves that the shot noise oscillates greatly in the region of $eV\geq \Delta_0$, while it increases monotonically in the region $eV<\Delta_0$. These behaviors are similar to the tunneling conductance. In tunneling through the G/SG/G-structure, transmission is enhanced by the active hole channels. As a result, the values of the shot noise are small in comparison with the Poisson value, which is $2eG$ corresponding to the uncorrelated transport. When the hole channels in addition to the electron channels contribute to the transport, the interference effect is strong and the shot noise is significantly suppressed. In the region of $eV<\Delta_0$, the impact of the proximity effect in the SG-region is strong giving rise to strong conductance and the values of $S$ and $F$ approach 0. In the case of $eV\geq \Delta_0$, the amplitude of oscillation decreases with the increase of $eV$; in the case of $eV<\Delta_0$, the amplitude of oscillation increases with the increase of $eV$, which originates from the same reason as the conductance.

\section{Conclusions}

Based on the Dirac--Bogoliubov--de Gennes equation and the scattering theory, we investigated the transport properties of the relativistic electrons and holes through the G/SG/G junction. We have deduced the analytical formulas of the tunneling conductance and the shot noise. Numerical results of the tunneling conductance and shot noise in the system are provided. We compared the two cases, one of which is in the presence of the specular Andreev reflection and the other of which is in the absence of the specular Andreev reflection. The physical results can be summarized as follows. Firstly, the conductance increases with the increase of the thickness of the SG-layer and the shot noise is suppressed by the conductance in the case of $eV=E_F$. Secondly, the potential strength significantly affects the transport properties. It suppresses the conductance and enhances the shot noise. Thirdly, we obtained similar results with the model of the G/SG junction in the condition $eV\leq \Delta_0$. In the limit of $a\rightarrow \infty$, the results of Beenakker for a G/SG junction can be reproduced. Fourthly, the thickness of the SG-layer affects the conductance and the shot noise more prominently in the condition $eV\geq \Delta_0$, causing the decrease of the oscillation amplitudes and the characteristic features of the specular Andreev reflection. In conclusion, the conductance is a combined result of the Andreev reflection and the specular Andreev reflection, which can be tuned by the system parameters; the shot noise is suppressed by the SG-barrier because of the contribution of the hole channels in addition to the electron channels.

\section{Acknowledgements}

This project was supported by the National Natural Science
Foundation of China (No. 11004063), and the Fundamental Research
Funds for the Central Universities, SCUT (No. 2014ZG0044).

\clearpage

\clearpage

\begin{figure}[h]
\includegraphics[height=7cm, width=10cm]{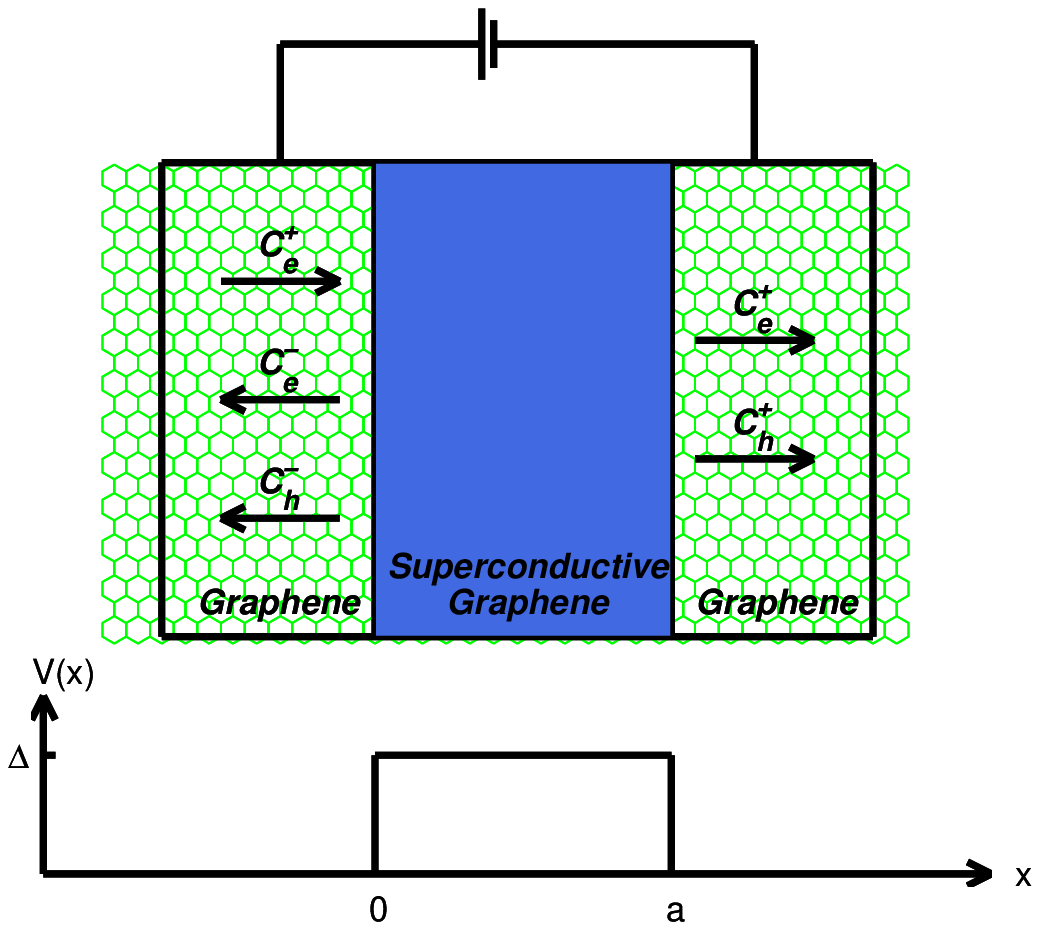}
\caption{Schematic illustration of reflection and transmission of quasipaticles in the graphene/superconductive graphene/graphene (G/SG/G) heterostructure, which occupies the $x$-$y$ plane. The left G region extends from $x=-\infty$ to $x=0$ and the right G region extends from $x=a$ to $x=+\infty$. The superconductive region occupies $0<x<a$. A voltage is applied between the left and right G regions. As a result of scattering by the SG, an incident electron can be transmitted and reflected both as an electron and as a hole into the right and left electrodes, respectively. The bottom is the superconducting potential profile with $a$ the SG-region width and $\Delta$ the value of the superconducting gap. }
\end{figure}

\clearpage

\begin{figure}[h]
\includegraphics[height=7cm, width=10cm]{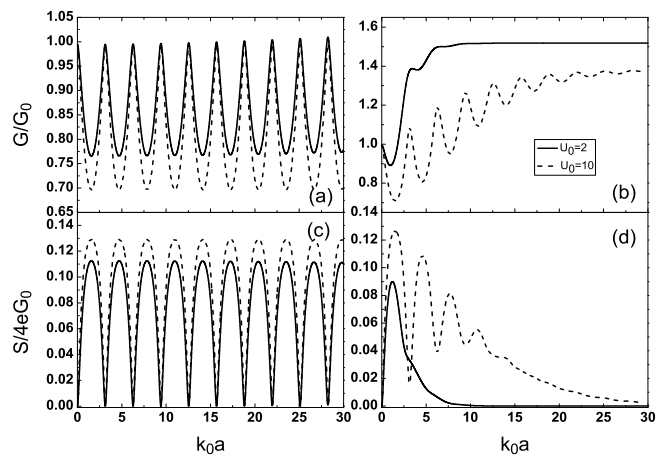}
\caption{Tunneling conductance and shot noise of the G/SG/G junction as a function of the SG-thickness $k_0 a$ for different potential strengths. Solid and dashed lines correspond to the potential strengths of $U_0/E_F=2 $ and 10, respectively. The other parameters are $\phi=0$, $\varepsilon/E_F=0.005$, and $a=5$. }
\end{figure}

\clearpage

\begin{figure}[h]
\includegraphics[height=10cm, width=11cm]{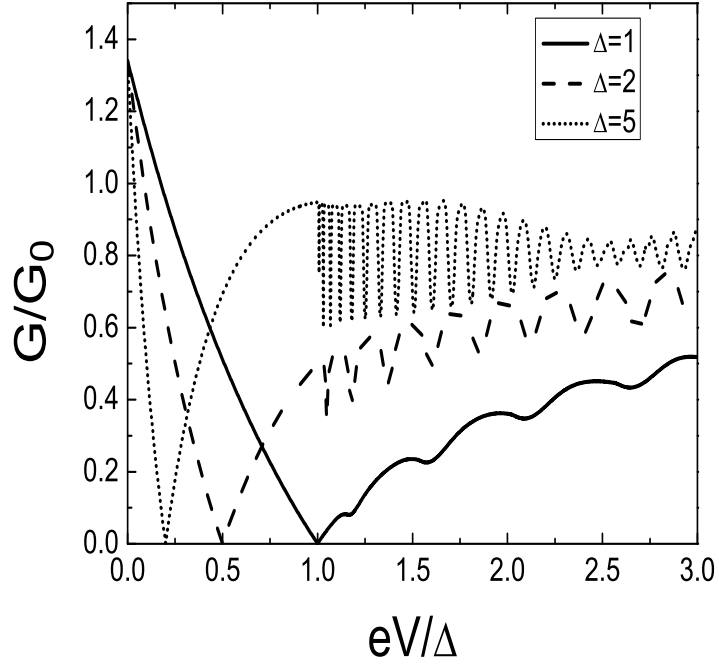}
\caption{ Tunneling conductance as a function of the bias voltage for different energy gaps of the SG. The solid, dashed, and dotted lines correspond to the energy gap in the SG region of $\Delta_0/E_F=1$, 2, and 5, respectively. The other parameters are $\phi=0$, $U_0/E_F=100$, and $a=5$.}
\end{figure}

\clearpage

\begin{figure}[h]
\includegraphics[height=10cm, width=12cm]{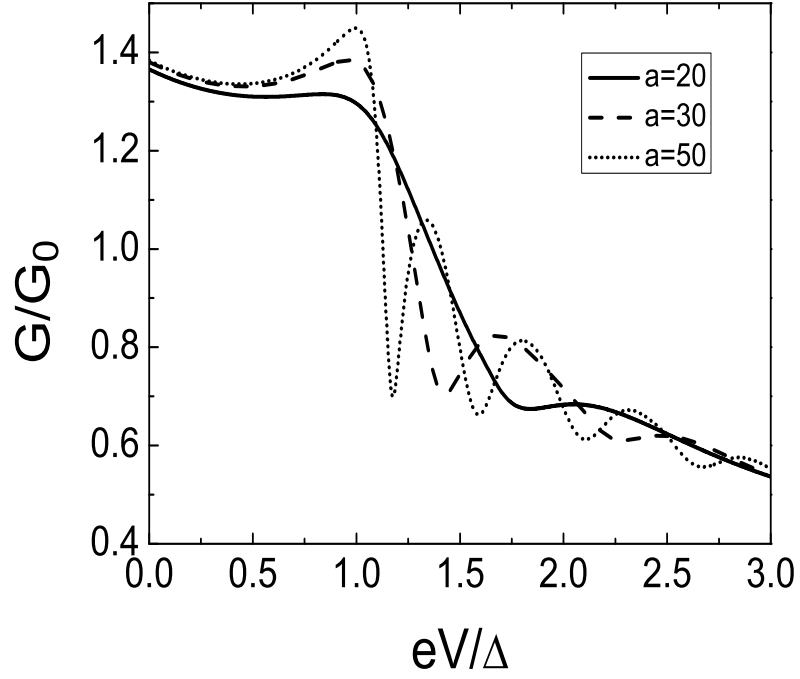}
\caption{Tunneling conductance as a function of the bias voltage for different thicknesses of the SG region. The solid, dashed, and dotted lines correspond to the SG-region thickness $a=20$, 30, and 50, respectively. The other parameters are $\phi=0$, $U_0/E_F=10$, and $\Delta_0/E_F=0.1$.}
\end{figure}

\clearpage

\begin{figure}[h]
\includegraphics[height=10cm, width=14cm]{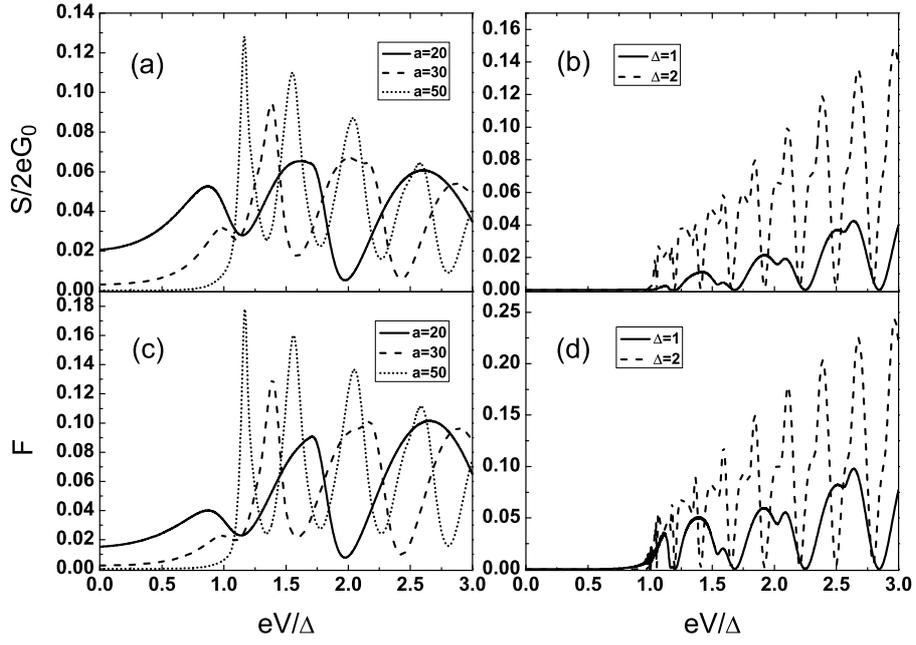}
\caption{Shot noise and the Fano factor as a function of the bias voltage for different energy gaps of the SG. The parameters in panels (a) and (c) are the same as Fig. 4. The parameters in panels (b) and (d) are the same as Fig. 3. }
\end{figure}

\clearpage

\end{document}